# Observation of Electronic Inhomogeneity and Charge Density Waves in a Bilayer La$_{2-2x}$Sr$_{1+2x}$Mn$_2$O$_7$ Single Crystal

Jeehoon Kim,[1] Junwei Huang,[1] J.-S. Zhou,[2] J. B. Goodenough,[2] J. F. Mitchell,[3] Alex de Lozanne[1,2]*

[1] Department of Physics, [2] Texas Materials Institute, University of Texas, Austin, TX 78712, USA

[3] Materials Science Division, Argonne National Laboratory, Argonne, Illinois 60439, USA

We employed a scanning tunneling microscope to image the (001) surface topography and local density of states (LDOS) in La$_{2-2x}$Sr$_{1+2x}$Mn$_2$O$_7$ (x=0.32, LSMO) single crystals below the Curie temperature ($T_C \approx 120$ K). The LDOS maps revealed a stripe-like modulation propagating along the tetragonal $a$ axis with a wavelength of about 16 Å, which is indicative of a charge density wave (CDW). The observed CDW in the x=0.32 sample is far from the Fermi surface nesting instability as compared with the data of angle resolved photoemission spectroscopy in an x=0.40 sample. The stripe model developed previously for cuprates can explain the observed CDW in our LSMO sample, indicating that competing interactions between localized and itinerant phases are the origin of the spatial modulations present intrinsically in cuprates and manganites.

Colossal magnetoresistance (CMR) is but one of several unusual phenomena found at the crossover from localized to itinerant phases in mixed-valent transition metal oxides (TMOs). The Ruddlesden-Popper (R-P) manganese-oxide phases have drawn much attention because the reduction of the 3-D perovskite structure to 2-D in these compounds amplifies the magnetic, electronic, and orbital fluctuations near the Curie temperature ($T_C$), thereby enhancing the CMR phenomenon in low magnetic fields [1]. Furthermore, as doping is increased from x=0.3 to x=0.5 the spins arrange themselves in antiferromagnetic, ferromagnetic, in-plane and out-of-plane configurations [2, 3]. In order to investigate the relationship between these competing degrees of freedom, intensive x-ray [4], neutron [5], optical conductivity [6], and angle-resolved photoemission spectroscopy (ARPES) [7-10] studies have been performed on the double-layered R-P manganites, predominantly on $La_{2-2x}Sr_{1+2x}Mn_2O_7$ with x = 0.40. Since it is possible to cleave the R-P manganites between two rock-salt planes to give an atomically flat perovskite surface, the $La_{2-2x}Sr_{1+2x}Mn_2O_7$ system lends itself to ARPES and STM measurements. The layered $La_{2-2x}Sr_{1+2x}Mn_2O_7$ system is, in particular, an ideal material to study the CMR effect for the ARPES experiment since in-plane momentum is conserved. Recent ARPES studies [7-10] on x = 0.36 and x = 0.40 samples have shown a well-defined Fermi surface and a pseudo gap at the Fermi energy $E_F$ below $T_C$, suggesting strong electron correlations. The shape of the Fermi surface indicates nesting, and therefore the possibility of a heretofore-unobserved CDW along a tetragonal $a$ axis. These ARPES experiments showed evidence for only a single itinerant-electron phase below $T_C$.

On the other hand, ARPES measurements [11] indicated a significant electronic density of states at the Fermi energy, originating from phase separation, in x = 0.36 and 0.38 samples above $T_C$. However, local metallic states in the insulating matrix above $T_C$ were not observed in the x = 0.4 sample in this experiment; magnetic fluctuations were observed instead. Strong and weak coupling scenarios in CMR physics have been debated for a long time. Localized charge stripes from the Jahn-Teller effect are a signature of strong electron-lattice coupling, while charge ordering based on CDWs can be explained by weak electron-phonon coupling [12]. Recently CDWs due to weak electron-phonon coupling were observed in the stripe phase of $La_{1-x}Ca_xMnO_3$ by optical-conductivity [13] and transport [14] measurements. A great deal of current research focuses on understanding the origin of exotic physics, such as stripes and checkerboard phases in high-temperature superconductors and CMR. A recent STM investigation [15] has revealed CDWs as the origin of the checkerboards in Bi-2201, though x-ray and neutron scattering experiments were not successful to observe CDWs possibly due to the weak and glassy nature of the charge modulations. The CDW observed in the superconducting state of Bi-2201 remains above the superconducting transition temperature, suggesting that a CDW could be responsible for the pseudogap phase in cuprates [15]. Since both manganites and cuprates share many aspects, including CDWs, the understanding of CDWs would play a salient role in revealing physics underlying exotic behaviors in these correlated systems. STM is a unique tool capable of probing charge modulations in real space with atomic resolution. The amount of STM data reported for the manganites is very small compared to those for the cuprates. Previously, a scanning tunneling spectroscopy (STS) measurement [16, 17] on an x = 0.30 sample showed a large energy gap from the strong confinement of polarons in a 2-D basal plane. In this Letter we report a CDW observed directly in a cleaved single crystal of $La_{1.36}Sr_{1.64}Mn_2O_7$ (x=0.32). To the best of our knowledge, this is the first real-space observation of a CDW in bilayered manganites. The CDW observed in the local density of states map can be better explained by the charge stripe model and is not consistent with a Fermi surface instability, as inferred from the doping dependence of its periodicity when comparing to the ARPES data in an x=0.40 sample. Our findings provide a venue to understand the origin and the role of CDWs in the CMR effect and the nature of an evolution of the electron-phonon interaction from the FS nesting to the charge stripe upon a decrease of the hole doping level.

Scanning tunneling microscopy/spectroscopy (STM/STS) measurements were made within a home-built ultrahigh vacuum low-temperature STM (UHV-LTSTM) system evacuated to a base pressure $P < 2 \times 10^{-10}$ Torr, while the pressure inside the cryogenic shield was at least an order of magnitude lower. We used chemically etched $Pt_{0.8}Ir_{0.2}$ tips, cleaned *in situ* by electron-beam bombardment. Oxygen-stoichiometric $La_{2-2x}Sr_{1+2x}Mn_2O_7$ (x = 0.32) single crystals ($T_C \approx 120$ K) were grown by the floating zone method in an IR image furnace [18]. The crystals were cleaved *in situ* immediately after they were taken out of the sample garage where they had been held against a cryogenic shield at 77 K in UHV, and they were immediately inserted into the LTSTM head. The $La_{2-2x}Sr_{1+2x}Mn_2O_7$ system has the tetragonal Ruddlesden-Popper $A_3M_2O_7$ structure [19, 20] in which an AO rock-salt layer is introduced periodically between layers containing two perovskite sheets. Cleavage takes place between two rock-salt planes, which allows imaging of the (001) surface topography of a perovskite layer [21]. The cleaved surfaces were atomically flat with a vertical corrugation of less than 1 Å. The exponential dependence of the tunneling current on tip-sample distance was repeatedly checked as a measure of the tunneling junction quality.

Figure 1(a) shows typical topographic images taken at 77 K (below $T_C \approx 120$ K). Randomly distributed small islands are ubiquitous. The average size of the islands is about one unit cell. In order to determine whether the bright spots are due to the presence of atoms adsorbed on the surface, we ran the STM images repeatedly over a month. If the bright regions were due to adatoms, the density of bright spots should increase significantly over the period of one month, but we found no change in the density of bright regions. Moreover, we found that bright spots fluctuate at 77 K: i.e., the two bright spots in the circles of Fig. 1(b) disappear in the next image, Fig. 1(c), and return in Fig. 1(d), which demonstrates that the bright spots are fluctuating at 77 K. However, the spots usually reappear in their previous location, which shows they tend to be trapped at a subtle lattice defect. In general, a thermally activated lateral motion of adsorbates at these temperatures is not expected on semiconductor or insulator surfaces [22]. Therefore, we can rule out the possibility that the bright regions represent surface adsorbate atoms. A lack of atomic resolution images has been reported in a previous STM study of this system, with the exception of a very small area [17]. A strong confinement of polarons in the basal plane was proposed as a possible explanation of the absence of atomic images in this system [17].

To investigate further the character of the clusters, we performed single point tunneling spectroscopy at 20 K at the dark and bright regions as indicated in the topographic images of Fig. 2(a). In order to rule out experimental artifacts, the same measurements were redone at the same locations, shown in Fig. 2(a), after a number of other measurements had been made. The following features are noted.

First, in the dark spots an energy gap of about 0.6 eV can be seen in the I-V curves, which is the same as the gap reported previously [17]. The magnitude of the energy gap in the bright spots is larger than that in the dark spots, indicating local inhomogeneities. Second, the I-V curves between bright and dark regions are strikingly different at high negative and positive bias, as shown in Fig. 2(b). Furthermore, the amplitudes of the fluctuations in the I-V curves of the bright regions are larger at high negative than at high positive bias, and these fluctuations occur in a range of bias voltage centered at about ±0.7 V. As reported in Refs. [10, 11], the underlying physics of the x = 0.4 sample is different from that of the x = 0.36, 0.38 samples. The phase separation scenario can be applied to x = 0.36 and 0.38 samples (low doping levels) whereas the magnetic fluctuation picture prevails in the x = 0.4 sample (optimal doping level). We believe that electronic fluctuations at the bright spots in the x = 0.32 sample are the result of a phase separation into conductive $Mn^{4+}$-rich clusters in an insulator $Mn^{3+}$ matrix. The conductive clusters have the σ-bonding 3d electrons of *e*-orbital symmetry delocalized into molecular orbitals of the cluster, but with $t^3$ configurations on each Mn giving a localized spin $S=3/2$. The $Mn^{3+}$ of the insulator matrix would have localized 3d electrons giving a spin $S=2$. The Mn-O bonds of the conductive clusters would be shorter than those of the matrix; the clusters would have a positive lattice charge and the matrix a negative lattice charge. At low temperatures, electrostatic and elastic forces would order the clusters into a lattice or CDW, but the charged regions of this CDW would not be static as in a conventional CDW; they would be coupled to the optical polarons propagating along the [100] and [010] Mn-O-Mn bond axes to provide the possibility of a cooperative mobility. The energy difference between the fluctuation peaks in Fig 2(b) is around 100 meV, which is close to the excitation energy of orbitals obtained by ARPES [10].

To investigate further the local electronic structure, scanning tunneling spectroscopy (STS) was performed at 20 K (well below $T_C \approx 120$ K). The differential tunneling conductance (*dI/dV*), which is proportional to the local density of states (LDOS), was recorded as a function of position (*x, y*) and energy (*E*) with a lock-in amplifier by adding a modulation voltage of 15 mV with a frequency of 718 Hz to the sample bias. The resulting three-dimensional matrix is shown as a differential conductance map and as a power-spectrum (PS) map in Fig. 3 and Fig. 4, respectively. The inhomogeneities are similar to those that have been found in high-temperature superconductors [23]. Figs. 3(a)-3(d) show the LDOS imaging of the filled states of the sample, and Figs. 3(e)-3(h) represent the imaging of the empty states of the sample. The stripe-like modulations were seen in all LDOS maps, particularly obvious at the negative bias voltage. This is our key observation. The stripe-like modulations in the filled states begin to show up above the energy gap and tend to organize well as the bias voltage increases. At negative sample bias (filled states of the sample), randomly distributed patches grow with increasing bias and coexist with the stripe modulations; however, the stripe modulations are still dominant. On the other hand, the randomly distributed patches appear around the energy gap and grow as the bias increases at positive sample bias; eventually they coexist with the same strength as the stripe modulation. The random patches in the conductance maps may be the result of inhomogeneities of $La^{3+}$ ions, since their DOS exhibit an unoccupied *f*-peak above the Fermi level [24].

These results are consistent with the single point spectroscopic data in Fig. 2(b). The degree of fluctuations at negative bias is more prominent than that at positive bias, which results in the clear observation of electronic modulation at the negative sample bias in the DOS maps.

To quantify the wavelength of the modulation, we obtained the Power spectrum (PS) of the LDOS maps, which is shown in Figs. 4(a)-4(d). The wavelength of the major modulation peak in the $E$ = -0.75 V map is ≈16 Å. After the STM measurement, we removed the sample from the vacuum chamber and used Laue x-ray back diffraction to obtain the direction of the modulation propagation in the tetragonal crystal plane; we confirmed it to be along the crystallographic *a* axis. At positive bias, small wave vectors around the center of the maps represent randomly distributed large islands. These small wave vectors are dominant at positive bias and coexist with the 16-Å modulation. The 16-Å modulation represents a charge density wave (CDW) that is prominent at negative bias voltage. Significantly, the wavelength of the 16-Å modulation does not change with bias voltage; it is non-dispersive, signaling quasiparticle interference scattering is not responsible for the observed CDW. Chuang *et al.* [7] used ARPES to determine the Fermi-surface (FS) topology of the hole-rich surfaces of x = 0.4 samples and on its basis predicted the formation of a CDW with a modulation of 3.3 lattice constants, *i.e.* ≈13 Å. Such modulation was subsequently observed by Campbell *et al.* [4] with x-ray diffuse scattering. Two types of CDW modulations are seen from the straight portions of the Fermi surface. Our STM data show that the system tends to form a CDW along the *a* or *b* crystal axes. In general, the wavelength of a CDW originating from a nesting instability of the FS shows a systematic doping dependence. The portion of the hole-like FS becomes smaller with decreasing hole doping, leading to an increase of the nesting wavevector; the greater the hole doping, the smaller the wavelength of the CDW. The wavelength of the x=0.32 sample (our STM result) is larger than that in the x=0.40 sample (ARPES measurement), which is inconsistent with the FS nesting scenario. Instead, the CDW observed in the x=0.32 sample may be explained by the stripe model [25] in which the wavelength of the stripe should decrease with increasing doping because more hole doping results in more stripes as a result of an energy balance between them. This scenario bears similarity to stripes reported previously in cuprates in the form of Wigner crystals where the periodicity of holes decreases with increasing doping [26]: especially, the charge stripe of the four-unit-cell periodicity (16 Å) in LSMO is similar to that found in cuprates with 1/8 hole doping [27]. The CDW is a macroscopic phenomenon that is not dependent on the presence of subtle inhomogeneities in the periodic potential, but it does reflect an increase in the electron-lattice interactions due to narrowing bandwidth. The CDW observed in the x=0.32 sample suggests the system tends to be more localized with decreasing doping, resulting in the charge stripe as a result of strong electron-phonon interactions, which is similar to what was reported in x=0.4 by neutron scattering [28]. Our findings suggest a further doping dependent study of CDW to investigate how the change of CDW properties affects CMR and MIT in conjunction with a drastic change of the spin structure.

The coexistence of paramagnetic insulating and ferromagnetic metallic states would be a ubiquitous picture [29], and together with lattice and charge degrees of freedom, the cooperative interactions take place between these competing phases, and thereby the formation of CDW in this system. Recent theoretical work suggested the aggregation of small polarons in bilayered manganese oxide sytems [30]. The aggregation of small polarons results in charge ordering with a periodicity of one or two unit cells whereas the charge modulation we observed is more than four unit cells, and is therefore too large for the small polaron ordering.

In conclusion, STM topography of the x=0.32 sample shows distinct islands indicative of a segregation into $Mn^{4+}$-rich, conductive clusters in an insulating $Mn^{3+}$-rich matrix. The conductive phase opens up channels for tunneling of electrons. Fluctuations in the I-V curves obtained with STM indicate the conductive islands are mobile along a tetragonal *a* axis. The mobility of the islands allows them to order at lower temperature into a lattice or CDW. STS conductance maps reveal a CDW that is not energy dispersive and a periodicity of the conductive clusters of about 16 Å. The CDW in the x=0.32 is also observed to be a global phenomenon. We point out that $Mn^{4+}$-rich clusters would have a smaller Mn-O equilibrium bond length than the $Mn^{3+}$ matrix, so elastic forces would limit the size of the clusters and couple their mobility to the optical polarons propagating along a tetragonal *a* axis. Moreover, since the $Mn^{4+}$-rich clusters would be coupled positive and the $Mn^{3+}$ matrix negative relative to the lattice, electrostatic forces would stabilize an ordering of the clusters into a lattice or a CDW. Moreover, such a CDW should be distinguished from a CDW instability arising from FS nesting since the conductive islands are independently mobile, i.e., capable of an order-disorder transition. We further point out that the observed CDW and dynamic phase-separation phenomenon reported have a striking similarity to that observed in superconducting cuprates. Our result indicates therefore that the spatial modulations in both these manganites and the cuprates are a general characteristic of $MO_2$ single or multiple perovskite sheets in a mixed-valent transition-metal system at the crossover from localized to itinerant electronic behavior. However, in the cuprates the hole-rich clusters are spin-polarized whereas in the manganites the hole-rich clusters, like the $Mn^{3+}$ matrix, carry a localized spin.

The authors thank A. Saxena, R. Movshovich, M. Graf, N. Haberkorn, and F. Ronning for the useful discussions. This work is supported by the National Science Foundation (DMR-0555663, DMR-0810119, DMR-1122603) and by the Welch Foundation.

* E-mail: delozanne@physics.utexas.edu

[1] J. F. Mitchell et al., J. Phys. Chem. B **105**, 10731 (2001).

[2] T. Asaka et al., Phys. Rev. Lett. **95**, 227204 (2005)

[3] J. Huang et al., Changbae Hyun, Tien-Ming Chuang, Jeehoon Kim, J. B. Goodenough, J.-S. Zhou, J. F. Mitchell, and Alex de Lozanne, Phys. Rev. B **77**, 024405 (2008)

[4] B. J. Campbell et al., Phys. Rev. B **65**, 014427 (2001).

[5] F. Moussa et al., Phys. Rev. Lett. **93**, 107202 (2004).

[6] H. J. Lee et al., Phys. Rev. B **62**, 11320 (2000).

[7] Y.-D. Chuang A. D. Gromko, D. S. Dessau, T. Kimura, Y. Tokura et al., Science **292**, 1509 (2001).

[8] N. Mannella et al., Nature **438**, 474 (2005).

[9] Z. Sun et al., Phys. Rev. Lett. **97**, 056401 (2006).

[10] Z. Sun et al., Proc. Natl. Acad. Sci. **108**, 11799 (2011).

[11] Z. Sun et al., Nature Physics **3**, 248 (2007).

[12] G. Subias et al., Phys. Rev. B **56**, 8183 (1997).

[13] A. Nucara et al., Phys. Rev. Lett. **101**, 066407 (2008).

[14] S. Cox et al., Nature Materials **7**, 25 (2008).

[15] W. D. Wise et al., Nature Physics **4**, 696 (2008).

[16] F. Massee et al., Nature Physics **7**, 978 (2011).

[17] H. M. Rønnow, Ch. Renner, G. Aeppli, T. Kimura, and Y. Tokura, Nature **440**, 1025 (2006).

[18] J.-S. Zhou, J. B. Goodenough, and J. F. Mitchell, Phys. Rev. B **61**, R9217 (2000).

[19] Y. Moritomo et al., Nature **380**, 141 (1996).

[20] T. Kimura, Y. Tomioka, H. Kuwahara, A. Asamitsu, M. Tamura, and Y. Tokura et al., Science **274**, 1698 (1996).

[21] F. Loviat et al., H. M. Rønnow, Ch. Renner, G. Aeppli, T. Kimura, and Y. Tokura, Nanotechnology **18**, 044020 (2007).

[22] G. A. Reider, U. Höfer, T. F. Heinz, Phys. Rev. Lett. **66**, 1994 (1991).

[23] S. H. Pan et al., Nature **413**, 282 (2001).

[24] G. W. Fernando, Handbook of Metal Physics **4**, 195 (2008).

[25] S. A. Kivelson et al., Rev. Mod. Phys. **75**, 1201 (2003).

[26] H. D. Chen, O. Vafek, A. Yazdani, and S. C. Zhang, Phys. Rev. Lett. **93**, 187002 (2004).

[27] V. J. Emergy, S. A. Kivelson, and J. M. Tranquada, Proc. Natl. Acad. Sci. **96**, 8814 (1999).

[28] F. Weber et al., Nature Mater. **8**, 798 (2009).

[29] S. Yunoki et al., J. Hu, A. L. Malvezzi, A. Moreo, N. Furukawa, and E. Dagotto, Phys. Rev. Lett. **80**, 845 (1998).

[30] C. Sen, G. Alvarez, and E. Dagotto, Phys. Rev. Lett. **98**, 1994 (2007).

FIGURE CAPTIONS

Fig. 1 (color online) (a) Typical topography in $La_{1.36}Sr_{1.64}Mn_2O_7$. (b)-(d) Topographic images sequentially taken one after another. Green circles represent the same place of the sample. All images were obtained at 77 K with the same tunneling condition: The set point voltage was $V_{set} = -1.0$ V and the set point current was $I_T = 100$ pA.

Fig. 2 (color online) Position-dependent tunneling spectroscopy in $La_{1.36}Sr_{1.64}Mn_2O_7$ at 20 K. (a) Topography ($I_T = 60$ pA and $V_T = -1.0$ V) with the marks at positions where the I-V measurements, shown in (b), were performed. Large fluctuations in the I-V curve, obtained in the bright region, were resolved at both negative and positive bias.

Fig. 3 (color online) Conductance (LDOS) maps above the energy gap ($|V| > 0.5$ V) of $La_{1.36}Sr_{1.64}Mn_2O_7$ at 20 K. Unprocessed LDOS maps in filled states (a)-(d) and empty states (e)-(h) of the sample. Stripe-like modulations are clear at the negative bias and coexist with large patches at the positive bias. The same color scale bar is used for all images (a)-(h).

Fig. 4 (color online) Power spectrum (PS) images obtained from the LDOS modulations in Fig 3. (a) PS of LDOS ($x, y, E = -0.75$ V) in Fig. 2. The modulation peak with a wavelength of 16 Å is clearly seen. (Inset) Schematic of PS; the white arrows represent Mn-O-Mn bonding directions. (b) PS ($x, y, E = -0.66$ V). The peak position does not change at $E = -0.66$ V: it is a non-dispersive peak. Small wave vectors around the center of the map show randomly distributed inhomogeneities in the map. (c)-(d) PS maps at the empty states of the sample. These PS maps also show non-dispersive behavior of the modulations, however, Fourier intensities of small wave-vectors are dominant.

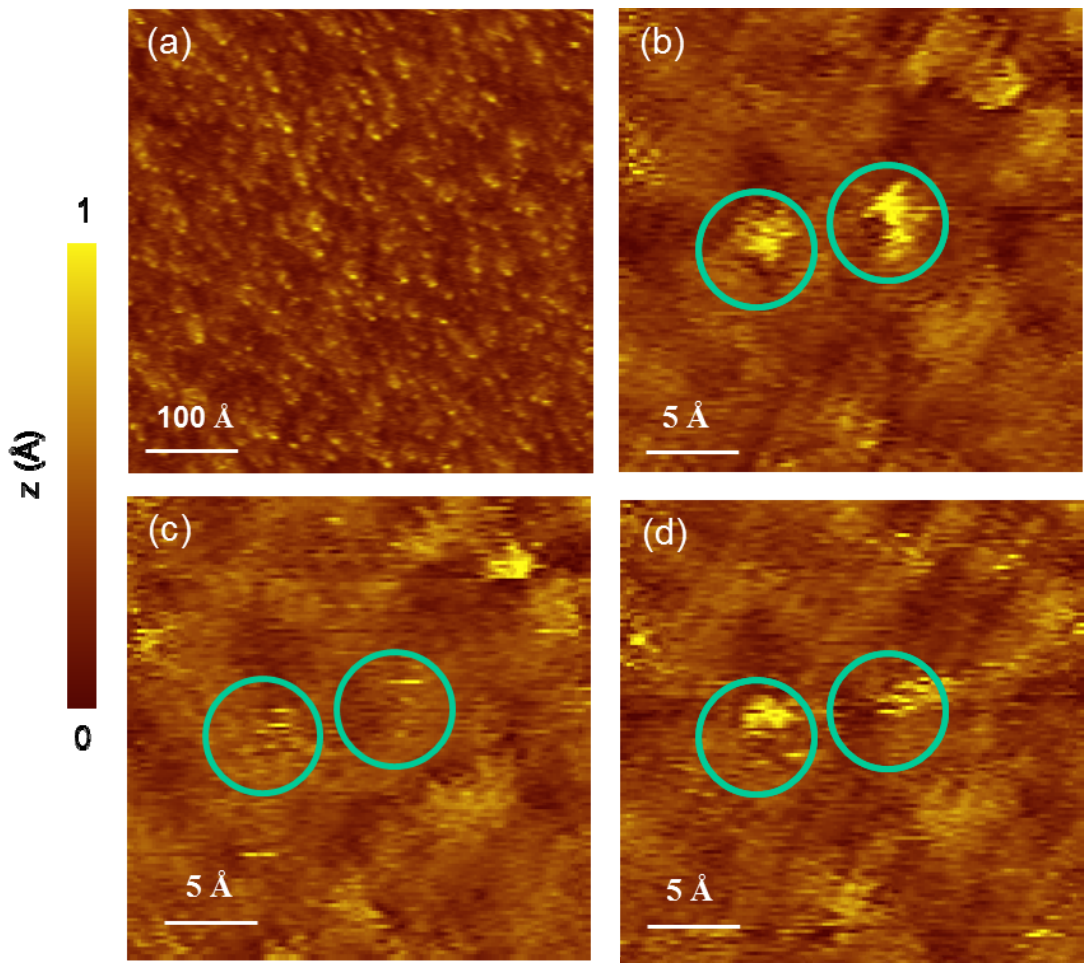

Fig. 1 (color online) (a) Typical topography in $La_{1.36}Sr_{1.64}Mn_2O_7$. (b)-(d) Topographic images sequentially taken one after another. Green circles represent the same place of the sample. All images were obtained at 77 K with the same tunneling condition: The set point voltage was $V_{set} = -1.0$ V and the set point current was $I_T = 100$ pA.

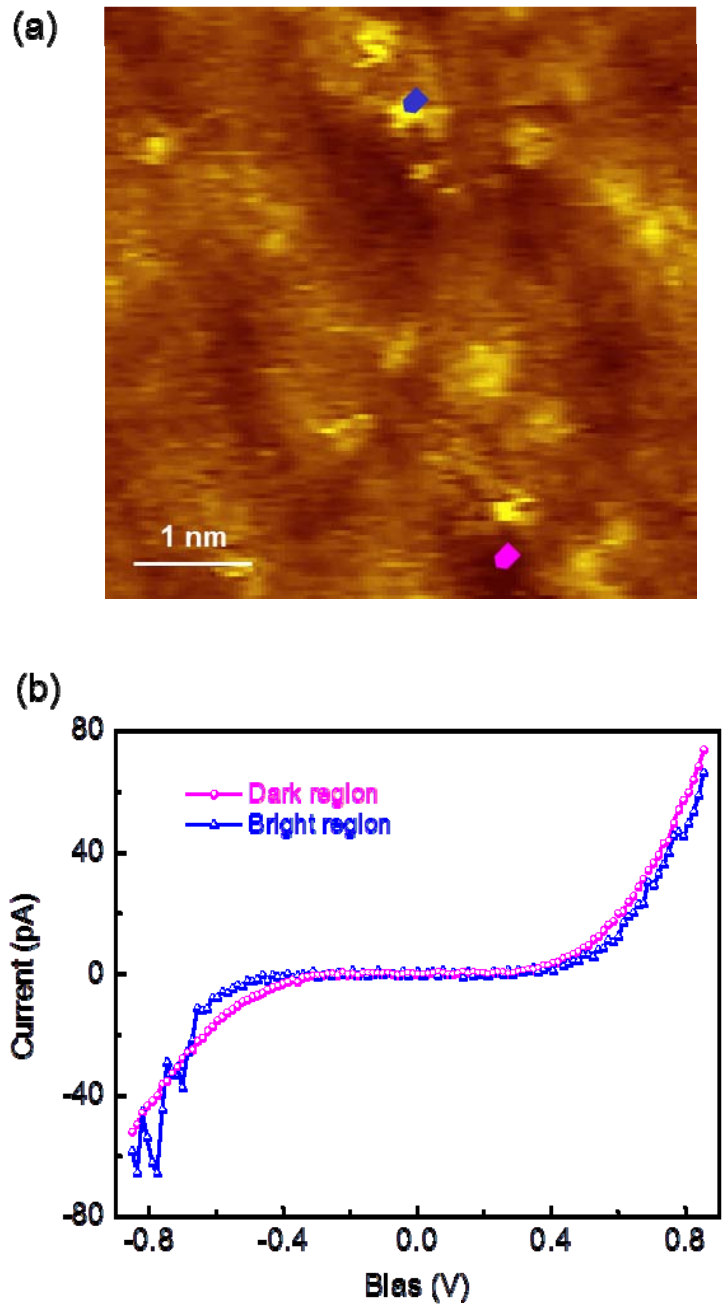

Fig. 2 (color online) Position-dependent tunneling spectroscopy in $La_{1.36}Sr_{1.64}Mn_2O_7$ at 20 K. (a) Topography ($I_T$ = 60 pA and $V_T$ = -1.0 V) with the marks at positions where the I-V measurements, shown in (b), were performed. Large fluctuations in the I-V curve, obtained in the bright region, were resolved at both negative and positive bias.

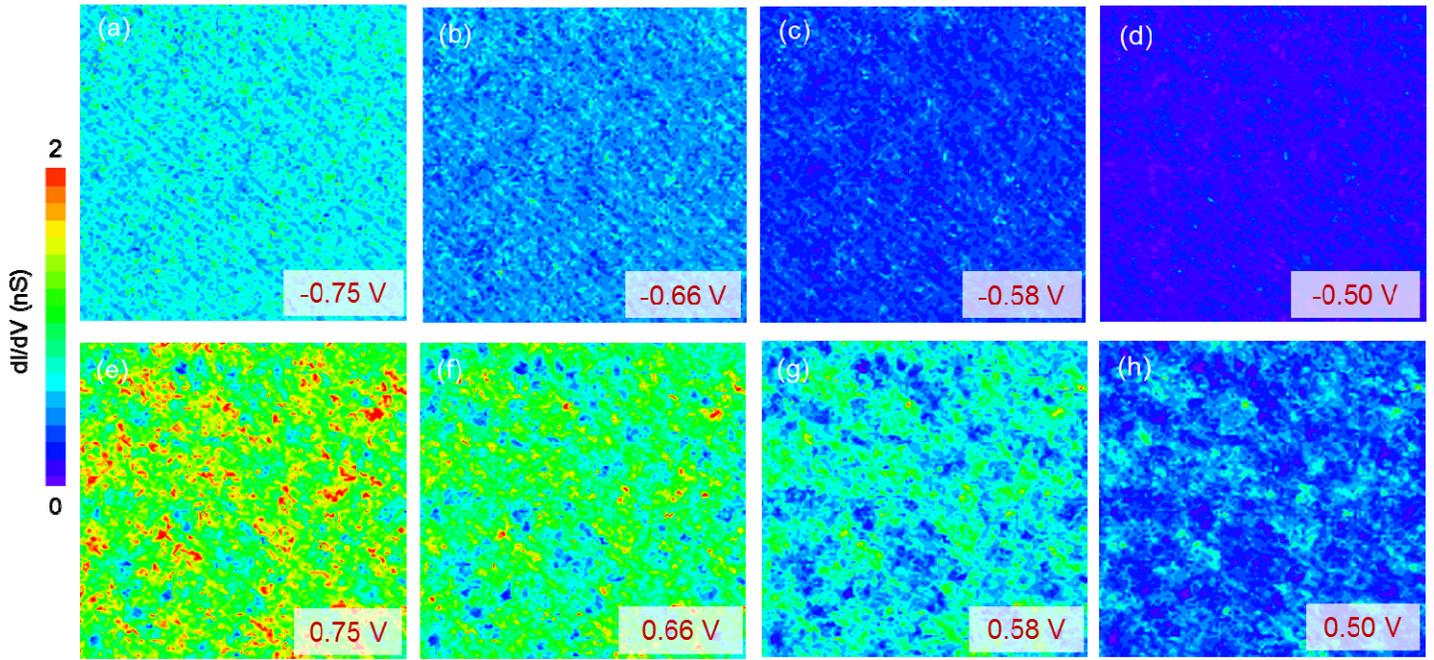

Fig. 3 (color online) Conductance (LDOS) maps above the energy gap ($|V| > 0.5$ V) of $La_{1.36}Sr_{1.64}Mn_2O_7$ at 20 K. Unprocessed LDOS maps in filled states (a)-(d) and empty states (e)-(h) of the sample. Stripe-like modulations are clear at the negative bias and coexist with large patches at the positive bias. The same color scale bar is used for all images (a)-(h).

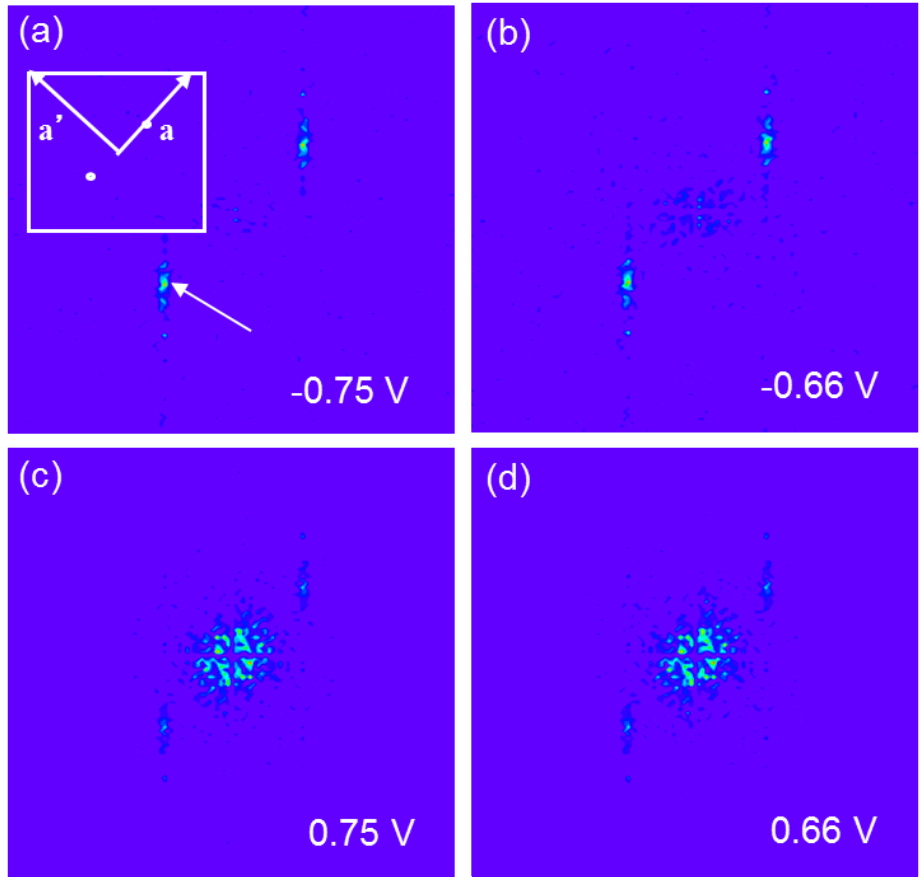

Fig. 4 (color online) Power spectrum (PS) images obtained from the LDOS modulations in Fig 3. (a) PS of LDOS ($x, y, E$ = -0.75 V) in Fig. 2. The modulation peak with a wavelength of 16 Å is clearly seen. (Inset) Schematic of PS; the white arrows represent Mn-O-Mn bonding directions. (b) PS ($x, y, E$ = -0.66 V). The peak position does not change at $E$ = -0.66 V: it is a non-dispersive peak. Small wave vectors around the center of the map show randomly distributed inhomogeneities in the map. (c)-(d) PS maps at the empty states of the sample. These PS maps also show non-dispersive behavior of the modulations, however, Fourier intensities of small wave-vectors are dominant.